%% file: paperPnotNP.tex
\documentclass[a4paper, twoside]{article}
\newtheorem{definition}{Definition}
\newtheorem{theorem}{Theorem}
\newtheorem{lemma}{Lemma}
\newtheorem{axiom}{Axiom}
\newtheorem{corollary}{Corollary}
\newtheorem{example}{Example}

\begin{document}

\title{$\, \mathcal { P } $ is not equal to $ \mathcal { NP }$.}
\author{Sten-{\AA}ke T{\"{a}}rnlund\thanks{ gmail: stenake.}
 \thanks{$\, 2^{nd} $ printing. \copyright \, Sten-{\AA}ke T{\"{a}}rnlund, 2009.}
\thanks{In the $ 2^{nd} $ printing the proof, in the $ 1^{st} $ printing, of theorem \ref{thsatp}  is divided into
three parts a new lemma \ref{lO}, a new corollary \ref{corH}, and the remaining part of the original
proof. 
There is no appendix in the $ 2^{nd} $ printing, but it is  available at www.arXiv.0810.5056v1. 
The  $\, 2^{nd} $ printing contains some simplifications, 
 more explanations, but no error has been corrected. The main results were first presented  6 August  2008 in a seminar at the department of Information Science of Uppsala University Sweden,  
 www.\,uu.\,se.}
}

\maketitle

\begin{abstract}

 $  \mathit   {SAT}   \not \in  {\cal P} $ 
is true, and provable in the simply consistent   extension $ {\mathbf  B}' $  of the first order 
 theory $ {\mathbf  B} $ of computing, with the single finite axiom  $   B $ characterizing  a universal Turing machine. Therefore $\, \mathcal { P } \neq \mathcal { NP } $ is true, and  provable in  the simply consistent extension   $\, \mathbf {B''} $ of theory $\, \mathbf {B} $. 
 \end{abstract}

\bibliographystyle{plain}

\input{introPnotNP}

\input{theoryPnotNP}

\input{computingPnotNP}

\input{CompPnotNP}

\subsection*{Acknowledgment}
Hanna-Nina Ekelund, Niklas Ekelund,  Andreas Hamfelt, Kaj B{\o}rge Hansen, Sophie Maisnier-Patin, Erik Nerep, J{\o}rgen Fischer Nilsson, Catuscia Palamidessi,
Torsten Palm, Alan Robinson, Thomas Sj{\"o}land, and Carl-Anton T{\"{a}}rnlund  thank you.\newline

\bibliography{references}
\end{document}

%% file: introPnotNP.tex
\section{Introduction}\label{secintro}
$ \, \mathit   {SAT}   \not \in  {\cal P} \,$ is true, theorem \ref{thsatp},  and provable, corollary \ref{corthsatnotinp}, in the simply consistent   extension $\, {\mathbf  B}' \,$  of  the first order theory $\, {\mathbf  B} \,$ of computing, with a single finite axiom
 $  B $  characterizing  a universal Turing machine.\footnote{Following  T\"arnlund  2008 \cite{Tarnlund:ioc}, a  simplification of T\"arnlund  1977 \cite{Tarnlund:HCC},  cf. Turing 1936  \cite{TUR:36} and Kleene 1967  \cite{KleeneSC:matl}. \label{fdesignB}}
  
  Therefore,  
$\, \mathcal { P } \neq \mathcal { NP } \,$ is true, theorem \ref{thpnp}, and provable,  corollary \ref{corthpnotnp},
in  the simply consistent extension   $\, \mathbf {B''} $ of theory 
$\, \mathbf {B} $, by the 
Cook-Levin theorem.\footnote{ $ \,\,\mathit   {SAT}     \in  {\cal P}\,\, \equiv \,\,\mathcal { P } = \mathcal { NP }, $   Cook 1971 \cite{Cook:71} and Levin 1973 \cite{Levin:usp}, cf. 
Sipser  2005 \cite{Sipser:comp}.
 \label{fcrt}}\,\,\footnote{  $\, \mathcal {P} $ vs $ \mathcal {NP} $  cf. Smale 1998 \cite{smale:98}, Cook 2003 \cite{cook:03}, and Cook $ www.claymath.org/millennium $.\label{fcpnp}}
$ \mathit   {SAT} $ is the set of satisfiable   propositional formulas.  $ {\cal P} $ is 
 the set  of classes of problems with solutions in polynomial computing time, for  a deterministic
Turing machine, in contrast,
$ {\cal NP} $   is the corresponding set  for a 
Turing machine.\footnote{cf. Karp 1972  \cite{Karp:pnp}, and 
Sipser  2005 \cite{Sipser:comp}.\label{fpnp}} 

The proof of theorem \ref{thsatp} 
is a proof   by contradiction  in the   extension 
$\, {\mathbf  B}' \,$  of   theory $\, {\mathbf  B}$, 
using the axiomatic method.\footnote{Hilbert and Bernays 1934 \cite{Hilbert&Bernays:gdm}, cf. Kleene 1967 \cite{KleeneSC:matl}. \label{faxiommeth}}\,\,\footnote{The axiomatic method is useful also in computing, 
e.g. proving that a program is correct, 
Clark and T\"arnlund  1977 \cite{Clark&Tarnlund:77},
and computability, T\"arnlund  2008 \cite{Tarnlund:ioc}.}
$\, {\mathbf  B}' $ is simply consistent\footnote{Simple consistency, cf. Kleene 1967 \cite{KleeneSC:matl}.\label{fsimplcon}} in the subset $  { \cal  U } $ of deterministic 
Turing machines that decide whether  a propositional formula is satisfiable or not,
corollary \ref{cor2defG}.\footnote{For the set of all
   Turing machines, however,   theory $ {\mathbf  B} $ is simply consistent if and only if  simple consistency of $ {\mathbf  B} $ is not provable in  $ {\mathbf  B} $, T\"arnlund  2008 \cite{Tarnlund:ioc}.\label{fsimplconB}} 
 This justifies the indirect proof method.

The  idea of the proof of theorem \ref{thsatp}, i.e. $ \,\mathit   {SAT}   \not \in  {\cal P} $, 
is  a relationship between computing time,\footnote{ The number moves of the tapehead  of a Turing machine, with an output, definition \ref{defhheadmoves},
 cf.  Hartmanis and Stearns 1965 \cite{hartmanis&stearns:hs65}, 
and Sipser  2005 \cite{Sipser:comp}. \label{fhssk}} 
 and proof complexity\footnote{ The size of a formal deduction (proof) in propositional logic,  
 cf. definition \ref{defsizeformded}.  \label{fCRpcomp}} in theory $\, {\mathbf  B} \,$. 
 
By  axiom $ B $, for a  Turing machine in  $  { \cal  U } $, which   decides 
 whether or not $ \neg F $ is satisfiable for 
 a sufficiently large tautology $ F $ $^{\ref{fsufflarge}}$ (on disjunctive normal form),
    there is a deduction in theory $ {\mathbf  B} $
    that is a propositional formula\,:
 \begin{equation}
 (Q_0 \supset F) \wedge (Q_{1} \supset Q_{0})\wedge \cdots \wedge (Q_{n} \supset Q_{n-1}) \wedge Q_n, \quad \label{eqYiFn}
 \end{equation}
 where   $ Q_0, \ldots, Q_n $ are atoms, and $ n $ 
 expresses the number of moves of the tapehead  of the Turing machine (computing time).

For several deduction systems including Robinson's resolution,\footnote{Robinson  1965 \cite{JAR:65}.\label{fRRP}}\,\,\footnote{The resolution system $ R $ is  used. This deduction system   is defined in 
(\ref{eqinfrileR}) - (\ref{eqprule1}).\label{fdsrR}} (\ref{eqYiFn}) yields a formal deduction
of $ F $,  i.e. 
\begin{equation} 
 Q_n, \neg Q_n \vee  Q_{n-1}, Q_{n-1},\ldots, Q_1, \neg Q_1 \vee  Q_{0}, Q_{0}, \neg Q_0  \vee F, F.
 \label{eqproofcomp}
 \end{equation}
Thus, (\ref{eqproofcomp}) is a proof of  $ F $  in  resolution system $ R $, i.e. $ \vdash_{_R} F $. Further,
the size of the resolution deduction of $ F $ is 
  the number of  symbols in  (\ref{eqproofcomp}),$^{\ref{fCRpcomp}}$ cf. lemma \ref{lO}.

Now, assume that  
\begin{equation}  { \mathit   {SAT} } \in { \cal P }. \label{eqsatinp} \end{equation}
Then, there is a Turing machine in $  { \cal  U } $ that decides whether  $ \neg F $ is satisfiable
or not in polynomial time in the size of $ F $, i.e. $ p(F) $.
Thus   \mbox{ for }(\ref{eqYiFn}) -  (\ref{eqproofcomp}),
\begin{equation}   n \leq p(F)\mbox{ for some } i \in { \cal  U } \mbox{ any sufficiently large tautology } F. \label{eqnleqpF} \end{equation} 
In addition (\ref{eqproofcomp}) gives,
\begin{equation} 
\label{eqsizeRded}
  \vdash_{_R} F \,\mbox{\, and \,  the size of } (\ref{eqproofcomp})  \leq p(F).    
\end{equation}
Now there is a contradiction by (\ref{eqsizeRded}) and the existence of sufficiently large tautologies  not
having a resolution proof in polynomial size, e.g. valid
pigeonhole formulas,\footnote{$ PHF_{n}^{m} $ for a propositional pigeonhole formula with  $ n $ holes and $ m $ pigeons, valid or not, where $ m \, n \in N $. The pigeonhole principle is: there is no injective function with a smaller co-domain than domain for finite sets. This idea can be expressed as a valid pigeonhole formula in disjunctive normal form, i.e. 
 $ PHP_{n}^{m} $ for $ \models PHF_{n}^{m} $ some $ m \, n \in N $. 
Some special cases of $ PHP_{n}^{m} $: $ m = n+1 $   is  classic (it also has the name $ PF_n $, cf. Cook and Reckhow 1979  \cite{cook&Reckhow:79}),  $ m \geq 2n $ is weak,  $ m = n^2 $ is very weak, and  $ m \to  \infty $ is still weaker. \label{fphf}}   Haken's theorem,\footnote{Haken's theorem 1985 \cite{haken85} for the classic pigeonhole tautology $ PF_n $:
every resolution proof of $ PF_n $ contains at least $ c^n $ different clauses for $ c > 1 $ some $ c \in R $ any sufficiently large $ n \in N $.\label{fhaken85}} and   Razborov's corollary.\footnote{A pigeonhole tautology $ PHP_{n}^{m} $, as a function, has a lower bound $ exp(\Omega(n^{1/3})) $   of the size of the proof in resolution, for an arbitrary 
$ m > n $. This is  Razborov's corollary 2003 \cite{wphpraz:03}.
\label{frazborov03}}

Thus,  assumption 
(\ref{eqsatinp}) is false, 
this explanation has arrived at theorem \ref{thsatp}\,:
\begin{equation}  { \mathit   {SAT} } \not\in { \cal P }. \label{eqsatninp} \end{equation}
 

Therfore  theorem \ref{thpnp}\,: $\, \mathcal { P } \neq \mathcal { NP } $, follows  by the 
Cook-Levin theorem.$^{\ref{fcrt}}$

It remains to present the crucial axiom $ B $, and the first order theory $ {\mathbf  B} $ of computing in section 
\ref{secfothcomp}.  
Notions of complexity are introduced in sections \ref{seccoc} and \ref{seccompHC}. Essentially, axiom $ B $  gives
  lemma \ref{lcomptimeYn2}, which yields lemma \ref{lO}. 
Then, theorem \ref{thsatp} is reached readily in the extension $ {\mathbf  B'} $ of  theory  $ {\mathbf  B} $.

%% file: theoryPnotNP.tex
\section{A theory  of computing }\label{secfothcomp}
The first  order theory $ \mathbf {B} $ of computing
 with a single finite axiom $ B $, which characterizes a universal Turing machine, is presented in this section.\footnote{The postulates of predicate calculus are employed in theory $ \mathbf {B}$. Frequently, $ G 4  $ of Kleene 1967 
\cite{KleeneSC:matl}, a Gentzen-type system    1934-5 \cite{GentzenG:untls} is used. 
 The  notation can simply be changed to a Hilbert-type system by Gentzen's theorem
 Kleene 1967 \cite{KleeneSC:matl}, and   to a resolution system by Robinson's theorem 1965 \cite{JAR:65}. Similarly, for    systems in Quine 1974 \S \, 37 \cite{quine:74}.\label{formal}} 
 Complexity   concepts  are introduced  in  sections \ref{seccoc} and \ref{seccompHC}.

Syntactically, there are two predicate symbols 
of theory $ \mathbf {B} $  written   $ T(i, a, u) $, and $ U(x, s, z, q,  j, i, u) $. In addition,   there is one function symbol $ \, . \, $ in infix notation $  x\,.\,y $ representing a list.  

A Turing machine  has a finite supply of arbitrary constant symbols, e.g.
the  alphabetic symbols, the natural numbers, and the symbols of propositional logic. For convenience, there is at least a subset of symbols, 
\begin{equation} \{ \emptyset, 0, 1,\,  _\sqcup   \} \subseteq K, \end{equation}
 where  $ K $ is a finite set of  constant  symbols,  and
$ _\sqcup  $  a blank symbol.

There are six sets for the Turing machines
in theory $ \mathbf {B} $.

 A finite set of states,\footnote{ $\, N $ is the set of the natural numbers. \label{fnatnumb}}
 \begin{equation}  Q \subset  N, \label{eqQ} \end{equation}
 where $ 0 $ is the halt state and $ 1 $ is the
start state. 

A finite set $ S $ of symbols,\footnote{ Operators\,: 
  $  \, \supset,  \, \equiv,  \, \vee,  \, \wedge, \, \neg,  \, \in, \,  =,  \, \leq, 
  \,\forall,  \,\exists,  \, \vdash, \,\models,\, \to  $, are ranked decreasingly to get simpler expressions. Thus,   $ \vdash B \to P \wedge  F   $  shall mean $  (\vdash (B \to P)) \wedge  F   $. Moreover, the  operators are used autonymously Kleene 1967 
\cite{KleeneSC:matl}.}  
\begin{equation}  S \,\mbox{ for } \, \{\,u : u \in K \, \vee \,
 (u = r\,. \,\emptyset \, \, \vee \,
 u = \emptyset\,.\,r) \, \, \wedge \, r \in K \} .\label{eqS} \end{equation}

The set $  D   $ of moves of  the tapehead of a Turing machine,  
 \begin{equation}   D    \,\mbox{ for } \,  \{   0, 1   \},\label{eqD}  \end{equation} 
 where $ 0 $ is a move to the left and $ 1 $ a move to the right.
 
There is a finite arbitrary large  two-way tape, with a left and right tape having an element  between them at the tapehead.\footnote{Turing 1936 \cite{TUR:36} has a one-way tape, for a two-way tape cf. Post 1947 \cite{post:rut}. \label{ftwtape} }
 Initially, the  two-way tape has an empty left tape, the input on the right tape,
 and the element between them has the symbol $ \emptyset $. When a computation starts the tapehead reads the symbol $ \emptyset $. The arbitrary long but finite two-way tape is represented as two lists.\footnote{Historically, Turing 1936 \cite{TUR:36} and  Kleene 1967  \cite{KleeneSC:matl} have
  a potentially infinite tape. In contrast, Davis 1958 \cite{DAV:58} and  Minsky 1967 \cite{Minsky:cfif} grow the arbitrary large finite tape in the computation. \label{fhisttape}} 
  
 The list on the right tape  grows to the right, if the symbol $ r\,. \,\emptyset \in S $  substitutes $ \emptyset $. The list on the left tape grows
to the left, if the symbol $ \emptyset\,. \,r \in S $  substitutes $ \emptyset $. 
Therefore the size of the two-way tape grows one element at a time 
controlled by the Turing machine.

 There are  two sets of lists for the two-way tapes,  $ L $  
for the right tapes, and $ L' $ for the left tapes.
 They are as follows.

The right tapes are lists of symbols,\footnote{\, $ \forall F $ for a free variable is universally
quantified over the entire formula $ F $.\label{funivquant}}
\begin{equation} L(\emptyset)
  \, \wedge\, \forall   \, \,( s\in S \wedge L(z) \supset L(s\,.\,z)). 
 \label{listL} \end{equation}
  
  The left tapes are lists of symbols,
  \begin{equation} L'(\emptyset) 
  \, \wedge\, \forall   \, \,( s\in S \wedge L'(z) \supset L'(z\,.\,s)). \label{listL'}\end{equation}
  
The set $ L $ of the  right tapes,
\begin{equation}  L   \mbox{ for }  \{u : \,\,
L(u)\,\}. \label{eqL} \end{equation}

The set  $ L' $ of the left tapes,
\begin{equation}  
L'   \mbox{ for }  \{u : \,\,
L'(u)\,\}. \label{eqL'} \end{equation}

 The set $ M  \subset L $ of codes of Turing
machines  is a subset of the set of lists.  The code of a Turing machine is a list of quintuples,\footnote{ cf. Turing 1936 \cite{TUR:36}.\label{ftmc}}
  \begin{equation} M(\emptyset)   \wedge \forall \,\,(p \,\, q  \in Q \wedge \, r \,\, s \in S 
 \wedge d\in D \wedge M(z) \supset M(p\,.\, s\,.\,q\,.\,r\,.\,d\,.\,z)).  \end{equation}

The set  $ M $ of codes of Turing machines,
\begin{equation}  
 M \mbox{ for } \{u : \,\,
 M(u)\,\}.\label{eqM}  \end{equation}
 
  The formulas of  $ \,\mathbf B $ are defined as usual in a first order theory.\footnote{cf. Kleene 1967  \cite{KleeneSC:matl}. \label{fdefformB}}

Semantically, the infix function symbol and the predicate symbols denote two 
functions\footnote{There are two  functions with similar syntax. (i) $\,.(r,z) \mapsto r\,.\,z $, where $ r \in S $ and $ z \in L $.
(ii) $\,.(x,r) \mapsto x\,.\,r $, where $ r \in S $ and $ x \in L'$. They are distinguished by their
appearance, function (i) on the right tape and (ii) on the left tape.\label{ffuncsymb}}  
and
two relations\footnote{ $ \, T \subseteq M \times L \times L' $, and 
$ U \subseteq L \times S \times L \times  Q \times M \times M \times L' $.}.
Of course, there is an intended  interpretation of the function symbol and the predicates in
the intended domains of
theory  $ \, \mathbf B $.
 
 $ T(i, a, u) $ shall mean  Turing machine $ i $  
 with input  $  a  $ computes an output  $ u $
 and  halts for $ i \in M \,\, a \in L \,\, 
 u \in L' $.\footnote{For example, the formula\,: $ A \supset B $, is written\,: 
 $ A\,.\,\supset\,.\, B \,.\,\emptyset $,
 as an input list. \label{finputlist}}\,\footnote{cf. Kleene p. 243 and footnote 167 Kleene 1967 \cite{KleeneSC:matl}.} 
 
 
 $ U(x, s, z, q,  j, i, u) $ shall mean Turing machine $ i $ computes $ u $ and  halts, where
$ x\,.\,s\,.\, z $ is the two-way tape of $ i $,$^{\ref{ftwtape}} $ 
$ s $ is a symbol (at the tapehead),   $ x $ is the left
tape, $ z $ is the right tape,  $ i $ is
in  state $ q $, and has an auxiliary code $ j $, for
$ i \,\, j \in M \,\, s \in S \,\, z \in L \,\, q \in Q \,\,
 x \,\, u \in L' $.

For instance, if $ T(i, A \, . \, \emptyset, \,\emptyset \, . \,\emptyset \, . \,  A \, . \, B) $  then
$ U(\emptyset, \emptyset, A \, . \, \emptyset \,. \, \emptyset, 1,  i, i, \,\emptyset \, . \,\emptyset \, . \,  A \, . \, B) $, 
i.e.  Turing machine $ i $ computes the output  
$ \emptyset \, . \,\emptyset \, . \,  A \, . \, B $.
It starts in state 1,  reads the symbol $ \emptyset $ (by its tapehead), and  the
 two-way tape  $ \emptyset, \emptyset, A \, . \, \emptyset \,. \, \emptyset $ is input. Here,  the left tape is the empty list $ \emptyset $, and the right tape is the list $ A \, . \, \emptyset \,. \, \emptyset $, where $ A \, . \, \emptyset  $ is the input list.\footnote{Note that the input list $\, A \, . \, \emptyset\, $ is  represented as
 $\, A \, . \, \emptyset\, .\,\emptyset $ on the right tape of  the two-way tape,
i.e. a list on a list. Thus, 
the beginning and end of  the input list are marked with the symbol 
$ \emptyset $. The beginning and end of the two-way tape itself  are also marked with  
$ \emptyset $. \label{ftwl}}
Thus, the output  added one element  $ B $ to the input list.

The axiom of theory $ \mathbf B $ has the name $ B $.$^{\ref{fdesignB}}$\,\footnote{The universal Turing machine in axiom 
$ B $ is also a logic program, i.e. 
for a Turing machine 
$ i \in M $ with input $ a \in L $ and output $ u \in L' $, $ B $ computes $ u  $, e.g. there is a deduction 
of $ u  $ in Prolog, cf. Kowalski 1974 \cite{KOW:74}, Colmerauer et al 1973 \cite{COLM:73}, 
and Warren 1977 \cite{Warren:al}. \label{fBalp}}
 It is  first written down, then
 an informal explanation of $ B $ follows.\footnote{The appendix of T{\"a}rnlund 2008 
  \cite{2008arXiv.5056T} shows examples of computations from axiom $ B $.  
 \label{exapp}}

\begin{axiom}\label{defB1}  
$ B $ for
\begin{eqnarray}
&& \forall \,\,T(i,\, a, \,u) \supset 
U( \emptyset, \, \emptyset,\, a \,.\,\emptyset, \,1,  \,i, \,i, \, u). \quad 
\label{defB101}\\
&& \forall\,\, U(\emptyset,\, \emptyset, \, a \,.\,\emptyset, \,1, \,i, \,i, \, u) \supset 
T(i,\, a,\, u).
\quad \label{defB10}\\
&& \forall \,\, U(x,\, s,\, z, \, 0, \,i, \, i,\, x).\quad \label{defB11}\\
&& \forall \,\, U( x,\, v, \,r\,.\, z  ,\, p, \, i, \,i, \,u) \supset
 U( x\,.\,v,\, s, \,z,\, q, \,q\,. \, s\,. \, p\,.\,r \,.\, 0\,.\, j, \,i, \,u).  \quad\label{defB12}\\
&&\forall \,\,  U(  x\,.\,r,\, v,\, z,\,p,  \,i, \,i, \,u) \supset
U( x,\, s,\,  v\,. \, z,\, q, \,q\,. \, s\,. \,  p\,. \, r\,. \,1\,. \, j,\, i, \,u). \label{defB13}\\
&& \forall \, \, U(x,\, s,\, z, \,q,  \,j, \,i,\, u) \supset
U(x,\, s,\, z, \,q,\, q'\,. \, s'\,. \, p\,. \,  r\,. \, d\,. \, j,\,i,\, u).  
\label{defB14}
\end{eqnarray}
Here,  $  \emptyset,\,\,  0 $ and $ 1 $ are constant symbols, and $  \,.\, $ a function symbol.$^{\ref{ffuncsymb}}$
\end{axiom}

The intended domains are the sets $ Q $ in (\ref{eqQ}),  $ S $ in (\ref{eqS}), $ D $ in (\ref{eqD}), $ L $ in (\ref{eqL}), $ L' $ in (\ref{eqL'}), and $ M $ in (\ref{eqM}).

 In sentences (\ref{defB101}) - (\ref{defB10}), there is an equivalence between  Turing machine 
 $ i $ with input $ a $ and output $ u $, and a universal Turing machine computing output $ u $ with the concrete representations of\,: the quintuples of $ i $,  the two-way tape holding input $ a $ and the symbol $ \emptyset $, the state $ 1 $, and the auxiliary code $ i $.

There is a halt  condition in sentence (\ref{defB11}), thus for state $ 0 $ the left  tape $ x $ 
is the output of the computation that stops.

 In sentence (\ref{defB12}),
if there is a new configuration (the tapehead  of Turing machine $ i $ has printed
$ r $ on the tape,  moved to the left, and entered state $ p $) then in the previous configuration  $ i $ is in state $ q  $ reading symbol $ s $ and has the  quintuple 
$ q\,.\,s\,.\,p\,.\,r\,.\,0 \,$. 

Similarly in sentence
(\ref{defB13}), 
if there is a new configuration (the tapehead  of  Turing machine $ i $ has printed
$ r $ on the tape, moved to the right, and entered state $ p $) then in the previous configuration $ i $ is in state $ q  $ reading symbol $ s $ and has the  quintuple 
$ q\,.\,s\,.\,p\,.\,r\,.\,1 \,$. 

In sentence (\ref{defB14}), Turing machine $ i $ is in state $ q $, reads  symbol $ s $, and 
searches for  a quintuple $ q\,.\,s\,.\,p\,.\,r\,.\,d \,$. 

Now, if   $ \exists u T(i, a, u) $ is true then $ \exists u T(i, a, u) $
is provable from 
 axiom \ref{defB1}, by induction on the number of moves of the tapehead in
  a proof.\footnote{The quantifiers some and any in the proof (meta) theory of $ \mathbf B $ are applied to the entire sentence in front of them. \label{fquantifiers}}\,\footnote{There is a proof of lemma \ref{ltp}  in T\"arnlund  2008 \cite{Tarnlund:ioc}.\label{fcons}}
\begin{lemma}\label{ltp}  
$ \,\exists u T(i, a, u) \, \supset \,\,\,  \vdash B \to  \exists u T(i, a, u) $
any $ i\in M $   $ a \in L $.
\end{lemma}

The notion of a Turing machine computation as  a formal proof in theory $ \mathbf B $,
is justified by axiom
\ref{defB1}, and lemma \ref{ltp}.\footnote{For a formal proof in  predicate calculus, cf. Kleene 1967  \cite{KleeneSC:matl}. 
\label{fformproofB}}\,\,${^{\ref{exapp}}}$
\begin{definition}\label{defcomp}  
A Turing machine  computation for  a formal proof in theory $ \mathbf B $.
\end{definition}

The set $ {\cal D } $ of deterministic Turing machines is  a proper subset of the set 
 of all Turing machines, and is introduced next. 
\begin{definition}\label{defdet}
 $ {\cal D }$  for $ \{ i : \, \,  i  $ has no
 quintuples $ q\,.\,s\,.\,p\,.\,r\,.\,d \,$ and  $\, q\,.\,s\,.\,p'\,.\,r'\,.\,d' $ and
   $\,\neg( p = p'\, \wedge \, r = r' \, \wedge \, d = d')  $ and $  i \in M  \} $. 
\end{definition}

%% file: computingPnotNP.tex
\section{Computing time }\label{seccoc}
In this section some complexity notions are introduced.

For a  Turing machine, with an output, the computing time is the number
of moves of its tapehead in the computation.$^{\ref{fhssk}}$ A Turing machine computation is
a formal proof in theory $ \mathbf B $, thus the computing time is the number of moves of the
tapehead in a proof in $ \mathbf B $.$^{\ref{formal}}$\,$^{\ref{fformproofB}}$\,\footnote{A move of the tapehead of a Turing machine   is specified in  (\ref{defB12}) - (\ref{defB13}) of axiom \ref{defB1}.\label{findpcs}}\,\footnote{Clearly, there is a relation that counts the number of
moves of the tapehead of a Turing machine in a proof in theory $ \mathbf B $, i.e.
$ H \subseteq E \times N $, where $ E $ is the set of proofs in theory $ \mathbf B $ and
$ N $ is the set of natural numbers. \label{frelH}}
 \begin{definition}\label{defhheadmoves}
 $ H(\vdash B \to  \exists u T(i, a, u), \, n) $ for 
 $ n $ is the number of moves of the tapehead of Turing machine $ i $ in a  proof of the sequent $  B \to \, \exists u T(i, a, u) \,$ in  $ G4 $
 any $ i  \in M $ $ a  \in L  \,\,\, n \in N$. 
 \end{definition}

The function $ \left|    \right| $  computes the size of an (input) list.\footnote{
 $\, \left|  \right|: L\to N $. \label{finputsize}}
 \begin{definition}\label{defno}
 $ \left| a \right| $ for the number of symbols of $ a\in L $.
\end{definition}

 A polynomial in the size of (the input) $ a $ is written  $  p( a ) $.

\begin{definition}\label{defin}
 $  p(  a ) $ for $   c \cdot | a |^q   $
 some $  c\,\, q \in N $ any $    a \in L   $.
 \end{definition}
 
The notion of a polynomial upper bound of the 
computing time is  introduced. 
 
\begin{definition}\label{defpolyupperboud}
 $ \vdash B \to  \exists u T(i, a, u) $ in $  p(a) $ for 
 $ H(\vdash B \to  \exists u T(i, a, u),\, n) \, \wedge \, n \leq     p( a )   $
  any $ i  \in M  \,\,   a \in L   $ some $ n \in N $.   
\end{definition}

\begin{definition}\label{defF}
  $ { \cal F } $ for the set of   formulas of propositional logic.
\end{definition}

The set $ { \mathit   {SAT} } $ of satisfiable   propositional formulas, and
the set $ { \mathit   {TAUT} } $ of tautologies are introduced.
\begin{definition}\label{defSAT}
 $ { \mathit   {SAT} } $ for $ \{ F \, :\,\,F  \in  {\cal F }  \, \wedge \, \not \models \neg F \}  $.
\end{definition}

\begin{definition}\label{defTAUT}
 $ { \mathit   {TAUT} }$ for $ \{ F \, :\,\, F  \in  {\cal F } \, \wedge \, \models F \, \}  $.
\end{definition}

 A name $ s $ is introduced for a  Turing machine that  computes whether 
a propositional formula$^{\ref{finputlist}}$ is  satisfiable, 
with output $\emptyset\,.\,0 $, or not  with 
 output $\emptyset\,.\,1 $.\footnote{Of course,   Turing machine $ s $ could compute the output satisfiable for $ \,\emptyset\,.\,0 $, 
 and unsatisfiable for  $ \emptyset\,.\,1 $. For reasons of space, the shorter version is used.
\label{flistrep}}
\begin{definition}\label{defh}
$\, T(s,\, a, u)\,  $ for $ \, a = F \,.\, \emptyset \,\,
\wedge \,\, ((u = \emptyset\,.\,0 \,\, \wedge \, \,\not\models  \neg F ) \,\,\vee \,\,
(u = \,\emptyset \,.\,1 \,\, \wedge \,\,\models \neg F)) \, $ $  s  \in  M  $ any $  a  \in  L \,\,\,F  \in  {\cal F } \,\,\,    u  \in  L $.
\end{definition} 
 
The set of correct deterministic Turing machines that decide whether a propositional formula
is satisfiable or not, is specified by
 Turing machine $ s $, by definitions \ref{defdet} and \ref{defh}.
 \begin{definition}\label{defS}
$ {\cal  U } $ for $ \{\, i \,: \, \exists u\, (T(i, \,a, u)\, \wedge
\, T(s,\,  a, u))\,\wedge  \, i  \in {\cal D } \,\, \wedge \,$ any $ a  \in L \,\}.$
\end{definition}

$ T(i, \neg F\,.\,\emptyset,\,\emptyset \,.\,1) $  if and only if $ F $ is a tautology for any 
 i in $ { \cal  U  } $ and  $ F $ in  $  { \cal   F } $.
\begin{corollary}\label{corlFaGH3}
$   T(i, \,\neg F\, . \, \emptyset, \,\emptyset \,.\,1) \, \equiv \,\,\,\models \, \to  F $
 any $ i \in  { \cal  U } \,\,\, F \in {\cal F }$.
  \end{corollary}
  
Theory $ \mathbf B $ is simply consistent in  $   {\cal U } $, i.e. there is no contradiction,
by axiom \ref{defB1}, lemma \ref{ltp}, and definitions \ref{defh} - \ref{defS}.$^{\ref{fsimplcon}}$\,\,$^{\ref{fsimplconB}}$

 \begin{corollary}\label{cor2defG}
 $ \vdash B \to (\exists u \, T(i,\, a,\,u ) \, \wedge \,
 \neg \exists u \, T(i,\, a,\,u )) $ no $ i \in  {\cal U } $ $  a \in L$.
\end{corollary}

In particular,
 $ \exists u\,T(i,\, a,\, u) $ is true if and only if  $  \exists u \, T(i,\, a,\, u)  $
 is provable from axiom $ B $ for $ i \in  {\cal U } \,\, a \in L$,
by axiom \ref{defB1}, corollary \ref{cor2defG}, lemma \ref{ltp}, and definition \ref{defS}.
 \begin{corollary}\label{cor3defG}
 $  \exists u\,T(i,\, a,\, u)\, \equiv \,\,\, \vdash B \to \exists u \, T(i,\, a,\, u)  $   
 any $ i \in  {\cal U } \,\, a \in L$.
\end{corollary}

The formula $    { \mathit   {SAT} } \in     { \cal P }  $ is introduced 
in the simply consistent  extension $ \mathbf B' $ of theory $ \mathbf B $,  corollary \ref{cor2defG}. Writing 
$ \,   { \mathit   {SAT} } \in     { \cal P }  $ for there is some  deterministic 
Turing machine  in $  { \cal U } $ that decides, in polynomial time,
whether any propositional formula 
is satisfiable or not, cf. footnotes \ref{fcrt} -
 \ref{fpnp}.
 
 \begin{definition}\label{defsatinp}
  $    { \mathit   {SAT} } \in { \cal P }  \,  $ for $ \, 
  \vdash B \to \exists u \,T(i, \, F\,.\,\emptyset, \, u) $ in  $ p( F )      $    
  some $ i \in  { \cal  U } $ any $ F \in { \cal   F } $.
 \end{definition}
 
 $ T(i, \neg F\,.\,\emptyset,\,\emptyset \,.\,1) $ is provable 
 from axiom $ B $ if and only if $ F $ is a tautology for any 
 i in $ { \cal  U  } $ and  
 $ F $ in  $  { \cal   F } $,
 by  corollaries \ref{corlFaGH3} and \ref{cor3defG}.
\begin{corollary}\label{corFa}
$ \vdash B \to  T(i, \, \neg F\,.\,\emptyset,\,\emptyset \,.\,1) \, \equiv \,\,\,\models \, \to  F $
 any $ i \in  { \cal  U } \,\,\, F \in {\cal F }$.
 \end{corollary} 
 
 If $    { \mathit   {SAT} } \in { \cal P }  \,  $ then 
 $ T(i,  \, \neg F\,.\,\emptyset, \,\emptyset \,.\,1) $ is provable from axiom $ B $, in polynomial time,
  for some deterministic Turing machine $ i $ in $ { \cal  U  } $ any tautology $ F $,
 by  definitions  \ref{defTAUT} and
\ref{defsatinp}, and
 corollary \ref{corFa}.

\begin{corollary}\label{corsatinp1}
 $    { \mathit   {SAT} } \in { \cal P }  \,  \supset \,\,\,
\vdash B \to T(i, \, \neg F\,.\,\emptyset, \,\emptyset \,.\,1) $ in  $ p( F )  \,
$    some $ i \in  { \cal  U } $ any $ F \in { \mathit   {TAUT} } $. 
  \end{corollary}

%% file: CompPnotNP.tex
\section{Computing time and proof complexity   }\label{seccompHC}
There is a relationship between  computing time\,$^{\ref{fhssk}}$  i.e., the number of moves of the tapehead of a Turing machine  in a computation, 
and proof complexity,$^{\ref{fCRpcomp}}$  i.e. the size of a proof in propositional logic.
The principal lemma \ref{lcomptimeYn2} gives this relationship  in lemma \ref{lO}, i.e.
 if a deterministic Turing machine in $   { \cal  U }  $ decides whether or not any sufficiently large negated tautology  is satisfiable in polynomial time then there is a proof of the tautology in polynomial size in resolution.
An example begins an explanation of this relationship.

\begin{example}\label{exV}
If a Turing machine in $   { \cal  U }\,\, $ decides whether or not any sufficiently large
negated tautology $ F  $ is satisfiable then the  sequent $ B \to  \,  T(i, \, \neg F\,.\,\emptyset, \, \,\emptyset \,.\,1) $ is provable for any $ i $ in $ { \cal  U }\,\,$ $
F $ in $ { \mathit   {TAUT} } $, in  Kleene's G4. Such a proof can be constructed, first,  by successive applications of the rule $ \forall\,\to $
in G4 getting the
 propositional conjunctions (\ref{defUn8}) to (\ref{defUn0})  from  axiom $ B $. Second, using
successive applications of the  rule $ \supset \,\to $ in G4. 
A  Turing machine in  $ { \cal  U } $ is deterministic, thus 
the  conjunctions (\ref{defUn0}) - (\ref{defUn8})  are   unique. 
$ \\ $

Writing $ A(i, \,  F, \, n) $  for  
 \begin{eqnarray}
 U(\emptyset\, .\, 1, s_{_{_{_{_{n}}}}}\!\!, z_{_{_{_{_{n}}}}}\!\!, 0,  i, i, \emptyset\, .\, 1) \wedge \quad\label{defUn0}\\
 {\Big(} U(\emptyset\, .\, 1, s_{_{_{_{_{n}}}}}\!\!, z_{_{_{_{_{n}}}}}\!\!, 0,  i, i, \emptyset\, .\, 1)  \supset
U(x_{_{_{_{_{n-1}}}}}\!\!, s_{_{_{_{_{n-1}}}}}\!\!, z_{_{_{_{_{n-1}}}}}\!\!, q_{_{_{_{_{n-1}}}}}\!\!,  j_{_{_{_{_{n-1}}}}}\!\!, i, \emptyset\, .\, 1) {\Big)}\wedge \cdots  \wedge
\quad\label{defUn1}\\
{\Big(}U(x_{_{_{_{_{m+3}}}}}\!\!, s_{_{_{_{_{m+3}}}}}\!\!, z_{_{_{_{_{m+3}}}}}\!\!, q_{_{_{_{_{m+3}}}}}\!\!, i, i, \emptyset\, .\, 1) \supset
U(x_{_{_{_{_{m+2}}}}}\!\!, s_{_{_{_{_{m+2}}}}}\!\!, z_{_{_{_{_{m+2}}}}}\!\!, q_{_{_{_{_{m+2}}}}}\!\!,  j_{_{_{_{_{m+2}}}}}\!\!, i, \emptyset\, .\, 1){\Big)} \wedge
 \quad\label{defUn200}\\
 {\Big(}U(x_{_{_{_{_{m+2}}}}}\!\!, s_{_{_{_{_{m+2}}}}}\!\!, z_{_{_{_{_{m+2}}}}}\!\!, q_{_{_{_{_{m+2}}}}}\!\!,  j_{_{_{_{_{m+2}}}}}\!\!, i,\emptyset\, .\, 1) \supset
U(x_{_{_{_{_{m+2}}}}}\!\!, s_{_{_{_{_{m+2}}}}}\!\!, z_{_{_{_{_{m+2}}}}}\!\!, q_{_{_{_{_{m+2}}}}}\!\!, i, i, \emptyset\, .\, 1){\Big)} \wedge  \cdots \wedge
 \quad\label{defUn3}\\
 \,\,{\Big(}U(x_{_{_{_{_{2}}}}}\!\!, s_{_{_{_{_{2}}}}}\!\!, z_{_{_{_{_{2}}}}}\!\!, q_{_{_{_{_{2}}}}}\!\!,  i, i, \emptyset\, .\, 1) \supset U(x_{_{_{_{_{1}}}}}\!\!, s_{_{_{_{_{1}}}}}\!\!, z_{_{_{_{_{1}}}}}\!\!, q_{_{_{_{_{1}}}}}\!\!, j_{_{_{_{_{1}}}}}\!\!, i,\emptyset\, .\, 1){\Big)} \wedge  \quad\label{defUn4}\\
{\Big(}U(x_{_{_{_{_{1}}}}}\!\!, s_{_{_{_{_{1}}}}}\!\!, z_{_{_{_{_{1}}}}}\!\!, q_{_{_{_{_{1}}}}}\!\!,  j_{_{_{_{_{1}}}}}\!\!, i, \emptyset\, .\, 1) \supset U(x_{_{_{_{_{1}}}}}\!\!, s_{_{_{_{_{1}}}}}\!\!, z_{_{_{_{_{1}}}}}\!\!, q_{_{_{_{_{1}}}}}\!\!,  i, i,\emptyset\, .\, 1){\Big)}\,\,\, 
\wedge \cdots \wedge \quad\label{defUn5}\\
{\Big(}U(x_{_{_{_{_{1}}}}}\!\!, s_{_{_{_{_{1}}}}}\!\!, z_{_{_{_{_{1}}}}}\!\!, q_{_{_{_{_{1}}}}}\!\!, i, i, \emptyset\, .\, 1) \supset U(\emptyset,\,\emptyset,\,  \neg F\, . \, \emptyset \, . \, \emptyset, 1,  j_{_{_{_{_{0}}}}}\!\!, i, \emptyset\, .\, 1){\Big)} \,\,\, \wedge \quad\label{defUn6}\\
{\Big(}U(\emptyset,\,\emptyset,\,  \neg F\, . \, \emptyset \, . \, \emptyset, 1,  j_{_{_{_{_{0}}}}}\!\!, i, \emptyset\, .\, 1) \supset U(\emptyset,\,\emptyset,\,  \neg F\, . \, \emptyset \, . \, \emptyset, 1,  i, i, \emptyset\, .\, 1){\Big)}
\wedge \cdots \wedge\quad\label{defUn7}\\
{\Big(}U(\emptyset, \,\emptyset,\, \neg F\, . \, \emptyset \, . \, \emptyset, 1,  i, i, \emptyset\, .\, 1) \supset 
T(i, \,  \neg F\,.\,\emptyset, \, \emptyset\, .\, 1){\Big)} \mbox { any }  i \in  { \cal  U }\,\,
F \in { \mathit   {TAUT} } \qquad \label{defUn8}\\
n \in N  \mbox { some }  \,\,  m \in N \,\,\, j_{_{_{_{_{m}}}}} \,\,\,j_{_{_{_{_{m+2}}}}}\, \, \,j_{_{_{_{_{n-1}}}}} \in { \cal  D } \,\,\,x_{_{_{_{_{m}}}}} \,\,\,x_{_{_{_{_{m+2}}}}}
\, \, \,x_{_{_{_{_{m+3}}}}}  \,\,\, x_{_{_{_{_{n-1}}}}} \in L' \qquad \label{defUn9}\nonumber\\
 z_{_{_{_{_{m}}}}} \,\,\,z_{_{_{_{_{m+2}}}}} \, \, \,z_{_{_{_{_{m+3}}}}}  \,\,\, z_{_{_{_{_{n-1}}}}} \in L\,\,\, s_{_{_{_{_{m}}}}} \,\,\,s_{_{_{_{_{m+2}}}}} \, \, \,s_{_{_{_{_{m+3}}}}}  \in S \,\,\,  q_{_{_{_{_{m}}}}} \,\,\,q_{_{_{_{_{m+2}}}}} \, \, \,q_{_{_{_{_{m+3}}}}} \in Q.\qquad \label{defUn10}\nonumber
\end{eqnarray}

Then,
\begin{equation}   
\vdash B \to T(i,\neg F\,.\emptyset, \emptyset\, .\, 1) \supset A(i, \,  F, \, n) 
\mbox{ any } i \in  { \cal  U }\,\,\, F \in { \mathit   {TAUT} } \mbox{ some } n \in N. 
\label{exA}\end{equation}

Not all of $ A(i, \,  F, \, n) $ in (\ref{exA}) is necessary, however,  to count the number
of  moves of the tapehead of a Turing machine  in the computation. 
Informally, writing $ V(i, \,  F, \, n) $ for (\ref{defUn0}) - 
(\ref{defUn3}), and (\ref{defUn8}).  This relation  
is a simplification of $ A(i, \,  F, \, n) $, but still
counts the number
of  moves of the tapehead of a deterministic Turing machine  in the computation.
\end{example}

In example \ref{exV}, the relation $ V(i, \,  F, \, n)$  counts the number of 
moves of the tapehead in a computation. It   is introduced more precisely as follows.
\begin{definition}\label{defVn}
$ V(i, F, n) \,\,$ for 
\begin{eqnarray}
\left(U(x_{_{_{_{_{0}}}}}\!\!, s_{_{_{_{_{0}}}}}\!\!, z_{_{_{_{_{0}}}}}\!\!, q_{_{_{_{_{0}}}}}\!\!,  i, i, \emptyset\, .\, 1)  \supset 
T(i,  \, \neg F\,.\,\emptyset,\,\emptyset \,.\,1)\right) \,\wedge 
\quad \label{pV10}\\
  U(\emptyset\, .\, 1, s_{_{_{_{_{n}}}}}\!\!, z_{_{_{_{_{n}}}}}\!\!, 0,  i, i, \emptyset\, .\, 1)\,\,\,\, \wedge \quad \label{pV11}\\
\bigwedge\limits_{{1 \leq \, m \, \leq n}} 
\left(U(x_{_{_{_{_{m}}}}}\!\!, s_{_{_{_{_{m}}}}}\!\!, z_{_{_{_{_{m}}}}}\!\!, q_{_{_{_{_{m}}}}}\!\!,  i, i, \emptyset\, .\, 1)  \supset 
U(x_{_{_{_{_{m-1}}}}}\!\!, s_{_{_{_{_{m-1}}}}}\!\!, z_{_{_{_{_{m-1}}}}}\!\!, q_{_{_{_{_{m-1}}}}}\!\!,  
  j_{_{_{_{_{m-1}}}}}\!\!, i, \emptyset\, .\, 1) \,\right) \, \wedge \quad
\label{pV12}\\
 \left(U(x_{_{_{_{_{m-1}}}}}\!\!, s_{_{_{_{_{m-1}}}}}\!\!, z_{_{_{_{_{m-1}}}}}\!\!, q_{_{_{_{_{m-1}}}}}\!\!,  
 j_{_{_{_{_{m-1}}}}}\!\!, i, \emptyset\, .\, 1)  \supset
 U(x_{_{_{_{_{m-1}}}}}\!\!, s_{_{_{_{_{m-1}}}}}\!\!, z_{_{_{_{_{m-1}}}}}\!\!, q_{_{_{_{_{m-1}}}}}\!\!, i, i, \emptyset\, .\, 1)\right)
\quad \nonumber\label{pV14}\\
 \mbox{ any } i  \in  { \cal  U } \,\,\, F \in { \mathit   {TAUT} }\,\,n \in N  \mbox{ some } 
   m \in N \,\,x_{_{_{_{_{m}}}}}\,\, x_{_{_{_{_{m-1}}}}} \in L' \,\,
 z_{_{_{_{_{n}}}}}\,z_{_{_{_{_{m}}}}} \,\, z_{_{_{_{_{m-1}}}}} \in L \,\,
 \quad 
 \label{pV16}\nonumber\\
 s_{_{_{_{_{n}}}}}\,\,s_{_{_{_{_{m}}}}}\,\, s_{_{_{_{_{m-1}}}}}\in S \,\,\,q_{_{_{_{_{m}}}}}\,\, q_{_{_{_{_{m-1}}}}}\in Q \,\, \, j_{_{_{_{_{m-1}}}}}\in { \cal  D }. 
 \quad 
  \label{pV17}\nonumber
\end{eqnarray}
\end{definition}

If a  Turing machine in  $ { \cal  U } $ decides whether or not $ \neg F $ is  satisfiable in computing time $ n $ 
then $ V(i, F, n) $ any 
$   F \in { \mathit   {TAUT} } $ $  n \in N $.
\begin{lemma}\label{lcomptimeVn}
 $ \,H( \vdash B \to T(i,  \, \neg F\,.\,\emptyset, \,\emptyset \,.\,1),\, n)  \,\, \supset  
 \,\, V(i, F, n) $ any $ i \in  { \cal  U } \,\,\, n \in N  $ $  F \in { \mathit   {TAUT}}$. 
  \end{lemma}
Proof.\footnote{Introduction of a star shall mean introduction of an assumption, 
and elimination of a star shall mean that an assumption is discharged, cf. Quine 1974 \S \, 37 \cite{quine:74}.\label{natded1}}
\begin{eqnarray}
\star & H(\vdash B \to T(i, \, \neg F\, . \, \emptyset, \,\emptyset \,.\,1),\, n) 
\quad i \in  { \cal  U }\,\,\,  F \in { \mathit   {TAUT}} \,\,\, n\in N \quad \label{plcomptimeVn10}\\
\star & \vdash B \to T(i, \, \neg F\, . \, \emptyset, \,\emptyset \,.\,1), \mbox{ definition } \ref{defhheadmoves}\quad \label{plcomptimeVn11}\\
\star &  V(i, F, n),\mbox{ axiom } \ref{defB1}, \mbox{ corollary } \ref{cor2defG},
\mbox{ and definition } \ref{defVn}  
\quad \label{plcomptimeVn13}\\
&  H(\vdash B \to T(i, \, \neg F\, . \, \emptyset, \,\emptyset \,.\,1),\, n) \,\, \supset \,\, 
 V(i, F, n) \mbox{ any } i \in  { \cal  U }\, \,\, F \in { \mathit   {TAUT}} \,\, n \in N
 \quad \label{plcomptimeVn14}
\end{eqnarray}

Atoms ("propositional variables") are usually introduced as names for large propositional formulas that are not only inconvenient, but also complex. For sufficiently large input,
 these atoms  give simpler propositional formulas 
 in a  standard syntax of propositional logic.

\begin{definition}\label{defQ_m}
$ Q_m $ for $  U(x_m, \, s_m, \, z_m, \, q_m,  \, i, \, i, \, \,\emptyset \,.\,1) $ 
any  $  i   \in  { \cal  U } \,\,\,x_m \in L' \,\,\,z_m \in L $ $ s_m  \in S \,\,\,  q_m \in Q \,\,\, m \in N $ some atom  $  Q_m \in { \cal F} $.
 \end{definition}

\begin{definition}\label{defR_m}
$ R_{m+1} $ for $  U(x_{m}, \, s_{m}, \, z_{m}, \, q_{m},  \, j_{m}, \, i, \,\emptyset \,.\,1 ) $  
any  $ i   \in  { \cal  U } \,\,\,j_{m}   \in { \cal  D } $ $ \,\,\,x_m \in L' \,\,\,z_m \in L \,\,\, s_{m}  \in S \,\,\,   q_{m} \in Q  \,\,\, m\,\in N $ some atom 
$ R_{m+1} \in { \cal F} $.     
 \end{definition}

Writing $ W(i, F, n) $ for an equivalent formula of $ V(i, F, n) $ any $ i \in { \cal  U } \,\,\, F \in { \mathit   {TAUT} } \,\,\, n \in N $,  by the auxiliary atoms  
in definitions \ref{defQ_m} - \ref{defR_m}.
\begin{definition}\label{defWn}
$ W(i, F, n) \,\,$ for $ Q_n \, \wedge \,\, \,\bigwedge\limits_{{1 \leq \, m \, \leq n}} (R_{m} \supset Q_{m-1})\, \wedge \,(Q_{m} \supset R_{m})\, \wedge \,  (Q_0 \supset T(i,\,\,  \neg F\, . \, \emptyset, \,\emptyset \,.\,1))$ any $ i  \in  { \cal  U } \,\,\, F \in { \mathit   {TAUT} }  \,\, n \in N $ some  $\,\, m \in N  \,  $ some atoms  
$\, Q_{0}, \ldots, Q_{n}, $ 
$R_{1}, \ldots, R_{n} \in { \cal F} $. 
\end{definition}

If $ V(i, F, n)\,$ then $ W(i, F, n) $ for  any 
$ i \in  { \cal  U } \,\,\, F \in 
 { \mathit   {TAUT} }  \,\,\, n \in N $,
by definitions  \ref{defVn} - \ref{defWn}. 
\begin{corollary}\label{cordefWn}
$ V(i, F, n)\, \supset W(i, F, n) $  any $ i \in  { \cal  U } \,\,\, F \in 
 { \mathit   {TAUT} }  \,\,\, n \in N $. 
  \end{corollary}

Writing $ Y(i, F, n) $ for  a simplified $ W(i, F, n) $.  
\begin{definition}\label{defYn}
$ Y(i, F, n) $ for $ (Q_0 \supset F)\,\wedge \, Q_n \, \wedge \, 
\bigwedge\limits_{{1 \leq \, m \, \leq n}} (Q_{m} \supset Q_{m-1})  $ 
any $ i  \in  { \cal  U } \,\,\, F \in { \mathit   {TAUT} }  \,\, n \in N $ some  $\,\, m \in N  \,$ some atoms $\, Q_{0}, \ldots, Q_{n}  $ $ \in { \cal F} $. 
 \end{definition}

If $ W(i, F, n)\,$ then $ Y(i, F, n) $ for  any 
$ i \in  { \cal  U } \,\,\, F \in 
 { \mathit   {TAUT} }  \,\,\, n \in N $,
by corollary \ref{corlFaGH3}  and  definitions \ref{defWn} - \ref{defYn}.

\begin{corollary}\label{corYn}
$ W(i, F, n)\, \supset Y(i, F, n) $  any $ i \in  { \cal  U } \,\,\, F \in 
 { \mathit   {TAUT} }  \,\,\, n \in N $. 
 \end{corollary}

If a  Turing machine in  $ { \cal  U } $  decides whether or not $ \neg F $ is  unsatisfiable 
in polynomial time then  
$ Y(i, F, n), $ where $ \, n \leq p(F),
$ any sufficiently large  $ F $ in ${ \mathit   {TAUT} } $ some 
$ n $ in $ N $.\footnote{A sentence $ C $ is true for sufficiently large natural numbers if
there exists a $ n' \in N $ such that the sentence $ C(n) $ is true for all $ n \ge n'$ $
n \in N $. \label{fsufflarge}}\,\footnote{ $ \lim\limits_{
{\left| F \right| \to \infty}\atop {F \in { \mathit   {TAUT}}}
} 
\frac {\left| U(x_m, \, s_m, \, z_m, \, q_m,  \, i, \, i, \, \,\emptyset \,.\,1)\right|}{\left| F \right|} = 1 $ and $ \lim\limits_{
{\left| F \right| \to \infty}\atop {F \in { \mathit   {TAUT}}}
} 
\frac {\left| U(x_m, \, s_m, \, z_m, \, q_m,  \, j, \, i, \, \,\emptyset \,.\,1)\right|}{\left| F \right|} = 1 $ for $ x_m \in  L' \,\,  s_m \in  S \,\, z_m  \in  L \,\,  q_m \in  Q \,\,  i \in  {\cal U }\,\,  j \in  {\cal D }\,\, m \in  N $,
  cf. definition \ref{defVn}.
\label{flim} }

\begin{lemma}\label{lcomptimeYn2}
 $ \vdash B \to T(i, \, \neg F\, . \, \emptyset, \,\emptyset \,.\,1) \mbox{ in }
 p(F)\,\,\supset \,\, Y(i, F, n) \, \wedge \, n \leq p(F) $ 
any  $  i \in  { \cal   U } $ any sufficiently large $ F \in { \mathit   {TAUT}} $ some $ n \in N $.
\end{lemma}
Proof.
\begin{eqnarray}
\star & \vdash B \to T(i, \, \neg F\, . \, \emptyset, \,\emptyset \,.\,1) \mbox{ in }
 p(F) \quad
 i \in  { \cal  U }\, \mbox{ sufficiently large } F \in { \mathit   {TAUT}}\quad \label{plcomptimeYn210}\\
\star & H(\vdash B \to T(i, \, \neg F\, . \, \emptyset, \,\emptyset \,.\,1),\, n) \,\,
\wedge \, n \leq p(F)
\quad\label{plcomptimeYn211}\\
\star &  Y(i, F, n) \mbox{ some }  n \in N , \mbox{ lemma  } \ref{lcomptimeVn}\, \mbox{  corollaries } \ref{cordefWn} -
 \ref{corYn} \mbox{ and footnote } \ref{flim}  \quad \label{plcomptimeYn212}\\
 &  \vdash B \to T(i, \, \neg F\, . \, \emptyset, \,\emptyset \,.\,1) \mbox{ in }
 p(F)  \,\, \supset \,\, 
 Y(i, F, n) \,\,\wedge \, n \leq p(F)
 \quad \label{plcomptimeYn215}\\
& \mbox{ any } i \in  { \cal  U }\,  \mbox{ any sufficiently large } F \in { \mathit   {TAUT}}
  \mbox{ some } n  \in  N \quad \label{plcomptimeYn216}\nonumber
\end{eqnarray}

\subsection{Proof complexity  }\label{subsubsecproofcomp}

Before the notion of the size of a deduction (proof) is taken up, the idea of a
formal deduction in propositional logic  is introduced.

The inference rule of the  Hilbert system $ H $ is modus ponens: 

\begin{equation} \frac{A,\,\,  A \supset B }{B } \label{eqmp} \end{equation}
where $  A $ and $  B $ are any propositional formulas.

In the Hilbert system $ H $, a formal deduction  is a finite list:
\begin{equation} F_{1}, \ldots, F_{n} \label{eqformded} \end{equation}
of formulas, where $ F_{k} $ is either an assumption formula $ A_{1}, \ldots, A_{m} $,
an axiom or follows from $ F_{i} $ and $ F_{j} $ by (\ref{eqmp}) for $ 1 \leq i \,\, j < k \leq n$. 
The  formal deduction (\ref{eqformded}) is a deduction in Hilbert system H of its last formula $ F_{n} $.\footnote{Following   Kleene 1967  \cite{KleeneSC:matl}. \label{fformded}}

In the Hilbert system $ H $, the existence of a formal deduction is written, 

 \begin{equation}
\Gamma  \vdash_{_{H}} F_{n}
\mbox{ for some formal deduction }  F_{1}, \ldots, F_{n}\mbox{ by } (\ref{eqmp})
\label{eqformdedH} \end{equation}
where $ \Gamma $ is a list of (zero or more) assumption formulas $ A_{1}, \ldots, A_{m} $. 
If there is no assumption, $ m < 1 $, then there is a formal proof
of $ F_{n} $, i.e. $ \vdash_{_{H}} F_{n}$.
 
 Hilbert systems are consistent and complete.\footnote{cf.  Kleene 1967 \cite{KleeneSC:matl}.}

Similarly,  for  Robinson's resolution systems,$^{\ref{fRRP} }$ which have a single inference rule.  
For the resolution  system $ R $ the inference rule is:

\begin{equation} \frac{A\vee P \vee B, \,\, \,\, C \vee \neg P \vee D }{A\vee B \vee C \vee D } \label{eqinfrileR} \end{equation}
 where $  A \,\, B \,\, C $ and $  D $ are any propositional formulas  on disjunctive normal form,
 DNF, (including the blank formula $ \sqcup $),
and  $ P $ is any propositional atom. 

A formal deduction in resolution is like a formal deduction in a  Hilbert system, 
except there is no logical axiom. In   resolution system $ R $, a formal deduction  is a finite list: 
\begin{equation} F_{1}, \ldots, F_{n} \label{eqformded1} \end{equation}
of formulas, 
where $ F_{k} $ is either an assumption formula $ A_{1}, \ldots, A_{m} $,
 or follows from $ F_{i} $ and $ F_{j} $ by (\ref{eqinfrileR}) for $ 1 \leq i \,\, j < k \leq n$. 
The list (\ref{eqformded1}) is a deduction in resolution system $ R $ of its last formula $ F_{n} $.

Hence, the existence of a formal deduction in  resolution system $ R $ is written,
\begin{equation} 
\Gamma  \vdash_{_{R}} F_{n} \mbox{ for there is a formal deduction }  F_{1}, \ldots, F_{n}  
\mbox{ by } (\ref{eqinfrileR}),\label{eqformdedR} \end{equation}
where $ \Gamma $ is a list of (zero or more) assumption formulas $ A_{1}, \ldots, A_{m} $,
if $ m < 1 $ then $ \vdash_{_{R}}  F_n $.

There is a proof rule for  indirect deductions (proofs).

\begin{equation} 
\mbox{ If } \,\Gamma, F' \vdash_{_{R}} \sqcup \, \mbox{ then } \Gamma\vdash_{_{R}}  F, 
 \label{eqprule1} \end{equation}
where $  F $ is a formula on disjunctive normal form, and $ F' \equiv \neg F $, where $ F' $ is
on conjunctive normal form (CNF), and $ \sqcup $ is the blank formula. 

If $ \Gamma $ is the empty list then there is a proof, 
i.e. if $ F' \vdash_{_{R}} \sqcup $ then $ \vdash_{_{R}}  F $.

Resolution systems are consistent and complete.\footnote{ 
If $  \Gamma \vdash_{_{R}} F $ then $ \Gamma \models F $ (consistency).
If $  \Gamma \models F $ then $ \Gamma \vdash_{_{R}} F $, where formulas  in
$ \Gamma $ are on CNF, and $ F $ on DNF (completeness), cf.
 Robinson's theorem 1965 \cite{JAR:65}. \label{frth}}

The size of a formal deduction $  F_{1}, \ldots, F_{n} $ is 
  the number of symbols in the formal deduction.
 Clearly, there is a relation,\footnote{$ G \subseteq K \times N $. $ K $ is the set of first order formal deductions, and $ N $ the set of the natural numbers. \label{frelG}} which computes the 
 size of a formal deduction in a Hilbert system, 
  and a Robinson resolution system.\footnote{$ \, \left| F \right| $ for the number of symbols of $ F  \in {\cal F }  $, cf. footnote \ref{finputsize} and
definition \ref{defno}. }
 \begin{definition}\label{defsizeformded}
$ G(\Gamma \vdash_{_{t }} F_{n},\, u) \,$ for 
$ u = \sum\limits_{{1 \leq \, k \, \leq n}}
\left| F_{k} \right| $ any $  u \in N $ some
formal deduction  $  F_{1}, \ldots, F_{n} $ in 
$ t $ with assumption $ \Gamma $, where $ t \in \{ H, 
 R \} $.
\end{definition}

A polynomial upper bound of the size of a  deduction is introduced.
\begin{definition}\label{defproofcompUnR}
$ \Gamma \vdash_{_{ t }} F $ in $ p( F ) $ for 
 $ G(\Gamma \vdash_{_{ t }} F, \, u) \,\wedge \, u   \leq   p(F) 
 $
 any $   F \in {\cal F } $ some $ u \in N $, where 
 $ t \in \{ H, 
  R \} $.   
\end{definition}

Then by lemma \ref{lcomptimeYn2} and the notion of a formal deduction in   resolution  system $ R $ the following
result follows.

If a Turing machine in  $ { \cal  U } $  decides whether or not $ \neg F $ is
satisfiable  in polynomial time then there is a polynomial deduction of $ F $ in
resolution system $ R $ for any sufficiently large tautology $  F $ on disjunctive normal form.   

\begin{lemma}\label{lO}
$ \vdash B \to T(i, \, \neg F\,.\,\emptyset, \,\emptyset \,.\,1) \mbox{ in  }  
p( F )    \supset \,\,\,
\vdash_{_{ R }} F $ in $ p( F) $ any $ i \in  { \cal  U } $ any sufficiently large  $ F \in { \mathit   {TAUT}} $ on disjunctive normal form.
\end{lemma}
Proof.
\begin{eqnarray}
\star & \vdash B \to T(i, \, \neg F\,.\,\emptyset, \,\emptyset \,.\,1) \mbox{ in  }  
 p( F ) \,\,  i \in  { \cal  U }  
\,\mbox{ a  sufficiently }  \quad \label{plO2}\\
& \,\mbox{ large } F \in { \mathit   {TAUT}} \mbox{ on DNF } 
 \quad \label{plO3}\nonumber\\
\star & (Q_0 \supset F)\, \wedge \, Q_{n} \,\wedge \,
\bigwedge_{{1 \leq \, k \, \leq n} \atop{n \leq p(F)}} (Q_{k} \supset Q_{k-1})  
\mbox{ some atoms } \quad \label{plO4}\\
&  Q_0,\,\ldots, Q_n 
  \in { \cal F} \mbox{ some } n \in N, \mbox{ lemma } \ref{lcomptimeYn2} \mbox{ and definition } \ref{defYn}\quad \label{plO5}\nonumber\\
\star & Q_{n}, \neg Q_{n} \vee Q_{n-1}, Q_{n-1}, \ldots, 
Q_{0}, \neg Q_{0} \vee F, F, \mbox{ a  deduction }     \quad \label{plO6}\\
&  \mbox{ in resolution system } R \mbox{ of  } F \mbox{ by } (\ref{eqinfrileR}) \mbox{ from } 
(\ref{plO4}) \quad \label{plO7}\nonumber\\
\star &  \vdash_{_{ R }} F  \mbox{ in  }   p( F ),
  \mbox{ definitions } \ref{defsizeformded}- \ref{defproofcompUnR}, \,\,
 (\ref{eqformdedR}) \mbox{ and  }(\ref{plO6})\quad \label{plO8}\\
 & \vdash B \to T(i, \, \neg F\,.\,\emptyset, \,\emptyset \,.\,1) \mbox{ in  }  
p( F )  \,\, \supset \,\,\, 
\vdash_{_{R}} F \mbox{ in } p( F) \mbox{ any  }  i \in  { \cal  U }
\mbox{ any  }\quad \label{plO9}\\
 &  \mbox{  sufficiently large } 
F \in { \mathit   {TAUT}} \mbox{ on DNF } \quad \label{pl10}\nonumber
\end{eqnarray}

There are  sufficiently large tautologies on disjunctive normal form   not having a proof in polynomial time in  resolution system R, e.g. valid pigeonhole formulas,$^{\ref{fphf}}$ cf. 
 Haken 1985 \cite{haken85},  and Razborov 2003 \cite{wphpraz:03}.$^{\ref{fhaken85}}$\,$^{\ref{frazborov03}}$ 

\begin{corollary}\label{corH}
$  \neg(\vdash_{_{R}} F 
$ in $ p( F ) $ any sufficiently large $ F \in { \mathit   {TAUT}} $
on disjunctive normal form).
\end{corollary}

 $ {\mathit {SAT}} $ is not in  $ \cal{P} $ in the simply consistent extension $ {\mathbf  B'} $
of theory $ {\mathbf  B} $,$ ^{\ref{frth}}$\,$^{\ref{fsufflarge}}$ by
  a proof by contradiction.

\begin{theorem}\label{thsatp}
$ {\mathit {SAT}} \not\in \cal{P} $ is true 
in the simply consistent extension  $ {\mathbf  B}' $
 of theory $ {\mathbf  B} $.
\end{theorem}
Proof.
\begin{eqnarray}
\star & {\mathit {SAT}}  \in {\cal{P}} \mbox{ in the simply consistent extension } {\mathbf  B}'
\mbox{ of  } {\mathbf  B}, \mbox{  definition } \ref{defsatinp} \quad \label{pthsatp11}\\
\star & \vdash B \to T(i, \, \neg F\,.\,\emptyset, \,\emptyset \,.\,1) \mbox{ in  }  
 p( F ) \,\,\mbox{ some }  i \in  { \cal  U } \mbox{ any }  \quad \label{pthsatp12}\\
 & \mbox{   sufficiently large } F \in { \mathit   {TAUT}}\mbox{ on DNF},\mbox{  corollary }\ref{corsatinp1} \quad \label{pthsatp13}\nonumber\\
\star &  \vdash_{_R} F  \mbox{ in }  p(  F ) \mbox{ any 
  sufficiently large  } F  \in { \mathit   {TAUT}} \,\mbox{ on DNF}, \mbox{  lemma } \ref{lO} 
  \quad \label{pthsatp14}\\
&  {\mathit {SAT}} \not\in {\cal{P} } \mbox{ in } {\mathbf  B}', \mbox{ contradiction 
} (\ref{pthsatp14}) 
 \mbox{ and corollary  } \ref{corH}, \mbox{ corollary  } \ref{cor2defG} \quad \label{pthsat15}
\end{eqnarray}
 
 In the extension $ {\mathbf  B''} $ of theory $ {\mathbf  B'} $
  the formula $ \cal{P} = \cal{NP} $  is defined    by the Cook-Levin theorem.$^ {\ref{fcrt}}$
 Then,  by theorem \ref{thsatp}.

\begin{theorem}\label{thpnp}
$\cal{P} \neq \cal{NP}$  is true 
in the simply consistent extension  $ {\mathbf  B''} $
 of theory $ {\mathbf  B} $.
\end{theorem} 

By lemma  \ref{ltp}, 
definitions  \ref{defS} - \ref{defsatinp}, and 
theorem \ref{thsatp}.

 \begin{corollary}\label{corthsatnotinp}
 $   {\mathit {SAT}} \not\in { \cal P } $ is  true, and
   provable in the simply consistent extension $ \mathbf {B}'$ of theory $ \mathbf {B}$.
 \end{corollary}

By lemma  \ref{ltp}, theorem \ref{thpnp}, and  corollary \ref{corthsatnotinp}.

\begin{corollary}\label{corthpnotnp}
 $ \mathcal { P } \neq \mathcal { NP } $ is true, and  provable in   
the simply consistent extension   $\, \mathbf {B''} $ of theory $\, \mathbf {B} $.
\end{corollary}